\documentclass[prc,aps,amssymb,showpacs,twocolumn]{revtex4}
\usepackage{graphicx}
\begin{document}

\title{Chiral Phase Structure at Finite Temperature and Density in Einstein Universe}
\author{Xuguang Huang , Xuewen Hao, and Pengfei Zhuang }
\affiliation{Physics Department, Tsinghua University, Beijing
             100084, China}
\begin{abstract}
The gravitational effect on the chiral phase structure at finite
temperature and density is investigated in the framework of
Nambu--Jona-Lasinio model in $D$-dimensional ultrastatic Einstein
universe. In mean field approximation, the thermodynamic potential
and the gap equation determining the curvature, temperature and
density dependence of the chiral condensate are analytically
derived. In the sense of chiral symmetry restoration and the order
of the phase transition, the scalar curvature of the space-time
plays similar role as the temperature.
\end{abstract}

\maketitle

\section {Introduction}

The phase structure of quantum field theory, especially  quantum
chromodynamics(QCD), was widely investigated in flat space-time in
the last two decades. It is quite interesting to make studies of how
the phase structure changes in the circumstance of very compact
stars and early universe where not only the temperature and density
effect but also the gravitational and gauge field effects can not be
neglected. The QCD phase transitions that probably happen in these
systems include mainly the restoration of spontaneously broken
chiral symmetry and the deconfinement from hadron gas to quark
matter and vice versa. They are closely related to the
hadrosynthesis at the early universe. We focus on the chiral phase
transition in this paper.

Four fermion models (and gauged four fermion models) are often
applied to describe the dynamical symmetry breaking in QCD,
electroweak theory and grand unified theory. One of the most
frequently used four fermion models is the Nambu--Jona-Lasinio (NJL)
model\cite{njl}. The chiral symmetry of this model is spontaneously
broken according to the emergence of a vacuum condensate of the
composite field of fermions, $\langle\bar{\psi}\psi\rangle\neq0$,
and the fermion mass is proportional to the condensate. The NJL
model is particularly convenient for the investigation of chiral
symmetry when some external conditions, like temperature, chemical
potential, non-trivial topology, external gauge field and gravity
field are taken into account. Such conditions need to be considered
in the study of high energy heavy ion collisions where the
temperature and baryon density are high enough, compact stars like
neutron stars and quark stars where the baryon density may be ten
times the normal nuclear matter density and the strong gravity and
external magnetic field can not be neglected, and early universe
where the gravity is certainly important and the temperature is
extremely high. Because of the non-perturbative phenomena around the
phase transition and the fermion sign problem in lattice QCD at
finite chemical potential, the NJL model has been widely considered
as a low energy effective theory of QCD in the study of chiral phase
transition in flat space-time\cite{njlreview}.

The chiral symmetry breaking in four fermion models in curved
space-time at zero temperature and chemical potential was
investigated in recent
years\cite{inag,inag2,inag3,ishi,eliz,eliz2,eliz3,corb}. It was
found that when the curvature of space-time is sufficiently large,
there will be no more spontaneous chiral symmetry breaking in the
vacuum. Therefore, in the sense of chiral symmetry restoration, the
curvature $R$ behaves like temperature $T$. This phenomenon can be
clearly seen in the discussion with both curvature and temperature
effects\cite{vita, inag2,inag4}. With the method of weak curvature
expansion for fermion propagator, the chiral phase structure of
Gross-Neveu\cite{gros} model in 2+1 dimensional space-time is
studied at finite chemical potential $\mu$ but zero
temperature\cite{kim}, the phase diagram in the $R-\mu$ plane is
similar to the diagram in the $T-\mu$ plane in flat space-time. In
the four dimensional curved NJL model at finite temperature and
chemical potential, the weak curvature approximation leads to a
first-order chiral phase transition\cite{goya}.

A natural question we ask is if we can use the method of weak
curvature expansion to treat the chiral properties around the
critical curvature $R_c$, when $R_c$ is not small. In this paper, we
study how the temperature, chemical potential and external gravity
influence the chiral symmetry of the NJL model in different
dimensions. We take the Einstein universe as the background
space-time, in which we can analytically calculate the thermodynamic
potential without using the weak curvature approximation in mean
field approximation. In flat space-time, the chemical potential
effect is very different from the temperature effect on the chiral
phase transition. For instance, the transition is of first order at
high chemical potential, while it is of second order at high
temperature. We want to know how the chemical potential affects the
chiral phase transition in highly curved space-time.

The paper is organized as follows. In Section \ref{s2}, we review
the NJL model in curved space-time at finite temperature and
chemical potential and obtain the thermodynamic potential in mean
field approximation. In Section \ref{s3} we analytically derive
the thermodynamic potential and gap equation for the chiral
condensate in the Einstein universe. The phase diagrams in
different planes are shown and discussed in Section \ref{s4}. We
summarize in Section \ref{s5}.

\section {The Curved NJL Model in Mean Field Approximation}
\label{s2}
In $D$-dimesional curved space-time with signature
$(+,-,-,-,\cdots)$, the action $S\left[\psi,\bar\psi\right]$ of
the $SU(N)$ NJL model takes the form of
\begin{equation}
\label{njl} S=\int d^D x
\sqrt{|g|}\left[\bar{\psi}i\gamma^{\mu}\tilde{\nabla}_{\mu}\psi
+\frac{G}{N}\left(\left(\bar{\psi}\psi\right)^2+\left(\bar{\psi}i\gamma^5\psi\right)^2\right)\right],
\end{equation}
where $g$ is the determinant of the metric tensor $g_{\mu\nu}(x)$,
$\psi(x)$ represents a fermion field in a $N$-dimensional internal
space, $\bar{\psi}=\psi^\dag\gamma^{\hat{0}}$ is the corresponding
Pauli conjugate spinor, the index with a hat denotes a standard
Dirac matrix defined in Minkowskian space-time, $G$ is the four
fermion coupling constant, $\gamma^\mu(x)$ is the Dirac matrix in
curved space-time satisfying $\{\gamma_\mu,
\gamma_\nu\}=g_{\mu\nu}$, $\tilde{\nabla}_\nu$ is the covariant
derivative with the chemical potential parameter $\mu$,
$\tilde{\nabla}_\nu=\nabla_\nu-i\mu\delta_{0\nu}=\partial_\nu+\Gamma_\nu-i\mu\delta_{0\nu}$
with $\Gamma_\nu(x)$ being the spinor
connection\cite{bril,park,naka,wein}, and $\gamma^5$ is defined as
the same as in flat space-time, it is independent of the
space-time coordinates. The model has a global $SU(N)\otimes
U(1)_L\otimes U(1)_R$ symmetry at $\mu=0$.

It is convenient to introduce the tetrads
formulism\cite{bril,park, naka, wein} to connect the definitions
in curved and flat space-times. Let $e_\mu^{\hat a}(x)$ be the
tetrads defined by $g_{\mu\nu}=\eta_{\hat a\hat b}e^{\hat a}_\mu
e^{\hat b}_\nu$, where $\eta_{\hat a\hat b}$ is the Minkowskian
metric, we can rewrite $\gamma^\mu=\gamma^{\hat a} e^\mu_{\hat
a}$, and derive $\Gamma_\mu=\frac{1}{8}e_\nu^{\hat a} e^{\nu \hat
b}_{\mu} \left[\gamma_{\hat a}, \gamma_{\hat b}\right]$ with
$e^{\nu}_{\mu \hat b} =\partial_\mu e^{\nu}_{\hat b}+\Gamma_{\mu
\rho}^\nu e^{\rho}_{\hat b}$ under the requirement $\nabla_\mu
\gamma^\nu=\nabla_\mu \gamma^5=0$, where $e_{\hat a}^\mu$ is the
inverse of $e^{\hat a}_\mu$.

The essential quantity characterizing a system in a grand
canonical ensemble can be taken to be the partition function $Z$.
It can be expressed in terms of the action of the system as
\begin{equation}
\label{z}
Z=\int\left[d\psi\right]\left[d\bar{\psi}\right]e^{iS\left[\psi,
\bar{\psi}\right]}.
\end{equation}
After bosonization, it becomes the integration over the boson
fields $\sigma=-2G\bar\psi\psi/N$ and $\pi=-2G\bar\psi
i\gamma_5\psi/N$,
\begin{equation}
\label{z2}
Z=\int\left[d\sigma\right]\left[d\pi\right]e^{iS_{eff}}
\end{equation}
with the effective action
\begin{eqnarray}
\label{effs}
S_{eff}\left[\sigma, \pi\right]&=&\int d^D x
\sqrt{|g|}\left[-\frac{N}{4G}\left(\sigma^2+\pi^2\right)\right]\nonumber\\
&&-i\ln
\mathrm{Det}\left[i\gamma^\mu\tilde{\nabla}_\mu-\left(\sigma+i\gamma^5\pi\right)\right],
\end{eqnarray}
where the determinant is taken in the internal space, the Dirac
space and the space-time manifold.

In flat space-time, in order to incorporate finite temperature
effects in quantum field theory, we can perform a Wick rotation
$t\rightarrow -i\tau$ and $\int dt\rightarrow -i\int_0^\beta d\tau$
with $\beta=1/T$, and the integration over fields is constrained so
that $\psi(0)=-\psi(\beta)$ and $ \bar{\psi}(0)=-\bar{\psi}(\beta)$
are satisfied for fermions. This is the so-called imaginary time
formalism of finite temperature field theory. However, in curved
space-time with non-static metric, the presence of time-space cross
terms makes it difficult to have a well-defined Wick rotation. Only
for the ultrastatic metric (for the static metric, we can
conformally transform it to the case of ultrastatic metric, see
refs.\cite{dowk,camp}), we can have a well-defined temperature by a
Wick rotation. The parameter $\mu$ has a clearly physical meaning of
chemical potential, since the number density is still conserved in
curved space-times, $\partial/\partial t\int
d^{D-1}\vec{x}\sqrt{|g|}\bar{\psi}\gamma^0\psi=0$. We will discuss
only the ultrastatic universe $\mathbb{R}\otimes\Sigma$ below with
$\Sigma$ being the $(D-1)$ dimensional space manifold.

With the known partition function, we obtain the thermodynamic
potential $\Omega$ as a function of temperature and chemical
potential,
\begin{equation}
\label{omega}
\Omega(T,\mu)=-\frac{T}{V}\ln Z(T,\mu),
\end{equation}
where $V=\int d^{D-1}\vec x\sqrt{|g|}$ is the volume of the space
manifold $\Sigma$.

In mean field approximation which is just the $O(1)$ order in the
large $N$ expansion of fermion propagator, the fields $\sigma$ and
$\pi$ are replaced by their thermal averages
$\langle\sigma\rangle$ and $\langle\pi\rangle$. To simplify the
notations, we express in the following the averages
$\langle\sigma\rangle$ and $\langle\pi\rangle$ by $\sigma$ and
$\pi$, respectively, without making confusion in the mean field
approximation. With the short notations, the thermodynamic
function can be simplified as
\begin{equation}
\label{omega2}
\Omega=\frac{N}{4G}\left(\sigma^2+\pi^2\right)-\frac{T}{V}\ln\mathrm{Det}\left[i\gamma^\mu\tilde{\nabla}_\mu
-\left(\sigma+i\gamma^5\pi\right)\right].
\end{equation}

If we do not consider the chiral anomaly induced by gravity, the
thermodynamic potential is invariant under chiral transformation,
and we can set $\pi=0$ without loss of generality. In this case, we
have
\begin{equation}
\label{omega3}
\Omega=\frac{N\sigma^2}{4G}-\frac{T}{V}\ln\mathrm{Det}\left[i\gamma^\mu\tilde{\nabla}_\mu-\sigma\right].
\end{equation}

The thermal expectation value $\sigma$ is the order parameter of the
chiral phase transition. The thermodynamic potential (\ref{omega3})
is a function of $T$ and $\mu$ with $\sigma$ initially an
undetermined parameter. In the spirit of thermodynamics, the
physical system is described only by $T$ and $\mu$. The order
parameter as a function of $T$ and $\mu$ is determined by the
minimum of the thermodynamic potential\cite{zhuang,hufn},
\begin{equation}
\label{min}
\frac{\partial\Omega}{\partial \sigma}=0,\ \ \ \ \
\frac{\partial^2\Omega}{\partial \sigma^2}\geq 0.
\end{equation}
Since we did not consider the initial fermion mass in the action
(\ref{njl}), the effective fermion mass produced through the
spontaneous chiral symmetry breaking is just the chiral
condensate, $m=\sigma$.

The measurable bulk quantities like pressure $p$, entropy density
$s$, fermion number density $n$, and energy density $\epsilon$ are
related to $\Omega$ by
\begin{eqnarray}
&& p=-\Omega,\ \ \ s=-{\partial\Omega\over \partial T}|_\mu,\ \ \
n=-{\partial\Omega\over \partial \mu}|_T,\nonumber\\
&& \epsilon=-p+Ts+\mu n.
\end{eqnarray}

Now we come back to the calculation of the mean field
thermodynamic potential (\ref{omega3}). In ultrastatic space-time
with line element $ds^2=dt^2-g_{ij}(\vec x)dx^idx^j$, it is easy
to show that the zero component $\Gamma_0$ of the spinor
connection is zero, and we have
$g^{\mu\nu}\nabla_\mu\nabla_\nu=\partial^2/\partial
t^2-\mathbf{\nabla}^2$ with
$\mathbf{\nabla}^2=g^{ij}\nabla_i\nabla_j$ being the spinor
Laplacian defined in the space manifold $\Sigma$. Taking into
account the relation
$\gamma^\mu\gamma^\nu\tilde{\nabla}_\mu\tilde{\nabla}_\nu=g^{\mu\nu}\nabla_\mu\nabla_\nu+\frac{1}{4}R-\mu^2-2i\mu\nabla_0$,
we have
\begin{eqnarray}
\label{det}
&&2\ln\mathrm{Det}\left[i\gamma^\mu\tilde{\nabla}_\mu-\sigma\right]\nonumber\\
&=&\ln\mathrm{Det}\left[i\gamma^\mu\tilde{\nabla}_\mu-\sigma\right]
+\ln\mathrm{Det}\left[\gamma^5(i\gamma^\mu\tilde{\nabla}_\mu-\sigma)\gamma^5\right]\nonumber\\
&=&\ln\mathrm{Det}\left[\gamma^\mu\gamma^\nu\tilde{\nabla}_\mu\tilde{\nabla}_\nu+\sigma^2\right]\nonumber\\
&=&\ln\mathrm{Det}\left[\left({\partial\over
\partial t}-i\mu\right)^2-\nabla^2+{R\over 4}+\sigma^2\right].
\end{eqnarray}
In order to complete the calculation, we should solve the
eigenequation of the operator in the last square bracket,
\begin{equation}
\label{eigen}
\left[\left(\frac{\partial}{\partial
t}-i\mu\right)^2-\nabla^2+\frac{R}{4}+\sigma^2\right]\Psi_k=\lambda_k\Psi_k,
\end{equation}
where $k$ stands for a set of complete quantum numbers describing
the eigenvalue $\lambda$ and eigenstate $\Psi$. By separating the
$t$ and $\vec x$ dependence of $\Psi$, the $\vec x$ dependent part
$\phi_l(\vec x)$ satisfies the corresponding eigenequation
\begin{equation}
\label{eigen2}
\left[-\nabla^2+\frac{R}{4}\right]\phi_l=\omega^2_l\phi_l
\end{equation}
in the space manifold and the normalization condition
\begin{equation}
\label{nor} \int d^{D-1}\vec x\sqrt{|g|}\phi_l(\vec
x)\phi_{l'}(\vec x)=\delta_{ll'}.
\end{equation}

If the solutions $\omega_l^2$ and $\phi_l(\vec x)$ of the
eigenequation (\ref{eigen2}) are known, we may take
\begin{eqnarray}
\label{eigen3}
&& \lambda_k=\omega^2_l+\sigma^2-(p_0-\mu)^2,\nonumber\\
&& \Psi_k(t,\vec x)= e^{ip_0t}\phi_l(\vec x),
\end{eqnarray}
where $p_0=i (2n+1)\pi/\beta$ is the fermion frequency with
$n=0,1,2,...$, since the eigenfunction $\Psi_k$ must be an
antiperiodic function of the period $\beta$.

After the frequency summation over $n$, we finally obtain the
thermodynamic potential in mean field
approximation\cite{habe,zhuang,hufn}
\begin{eqnarray}
\label{omega4}
\Omega&=&\frac{N\sigma^2}{4G}-\frac{N}{V}\sum_ld_l\Big[E_l
+T\ln\left(1+e^{-{E_l-\mu\over
T}}\right)\nonumber\\
&&+T\ln\left(1+e^{-{E_l+\mu\over T}}\right)\Big],
\end{eqnarray}
where $E_l=\sqrt{\omega^2_l+\sigma^2}$ is the quasiparticle energy
and $d_l$ is the degeneracy of the $l$-th eigenvalue of equation
(\ref{eigen2}).

\section {The Gap Equation in the Einstein Universe}
\label{s3}
The question left is to solve the eigenequation (\ref{eigen2}). To
this end, some approximate methods like weak field
approximation\cite{inag,inag3,kim,goya} and high temperature
expansion\cite{camp,camp2} are widely used. However, around the
phase transition, the interaction among the constituents of the
system should be very strong and those methods based on perturbative
expansion are in principle not suitable for the study of the phase
structure. In fact, the eigenequation (\ref{eigen2}) can be exactly
solved in some special universes\cite{camp2,inag3,cand}, for
example, the static Einstein universe. The $D$-dimensional Einstein
universe is represented by the line element
\begin{equation}
\label{einstein}
ds^2=dt^2-a^2(d\theta^2+\sin^2\theta
d\Omega_{D-2})
\end{equation}
defined with the topology $\mathbb R \otimes \mathbb S^{D-1}$,
where $a$ is related to the curvature, $R=(D-1)(D-2)a^{-2}$. With
the eigenvalues\cite{cand,camp2} of the spinor Laplacian on the
sphere $\mathbb S^{D-1}$, the degeneracy $d_l$, the eigenvalue
$\omega_l$ and the volume $V$ can be expressed in terms of $l$ and
$R$ as,
\begin{eqnarray}
\label{einstein2}
&& d_l={2^{[(D+1)/2]}\Gamma(l+D-1)\over
l!\Gamma(D-1)},\ \ \ \ \ \omega^2_l(R)=b_l^2 R,\nonumber\\
&&b_l^2=\frac{(2l+D-1)^2}{4(D-1)(D-2)}+\frac{1}{4},\ \ \ \ \ l=0,1,2,\cdots,\nonumber\\
&& V(R)={2\pi^{D/2}a^{D-1}\over \Gamma(D/2)},
\end{eqnarray}
where $[x]$ is the floor function of $x$, and the summation over
the eigenvalues in the mean field thermodynamic potential becomes
now explicit,
\begin{eqnarray}
\label{omega5}
\Omega&=&\frac{N\sigma^2}{4G}-\frac{N}{V}\sum_{l=0}^\infty d_l
e^{-{\omega_l\over \Lambda}}\Big[E_l+T\ln\left(1+e^{-{E_l-\mu\over
T}}\right)\nonumber\\
&& +T\ln\left(1+e^{-{E_l+\mu\over T}}\right)\Big].
\end{eqnarray}
Because of the contact interaction among the particles, the NJL
model is non-renormalizable when $D\geq4$, and in general it is
necessary to introduce a regulator that serves as a length scale
in the problem. That is the reason why we have introduced a soft
cutoff factor $e^{-\omega_l/\Lambda}$ in the summation over $l$.
Otherwise, the first term in the square bracket which is the
contribution from the vacuum will be divergent.

There are two parameters in the model, the coupling constant $G$ and
the soft cutoff $\Lambda$. In the case of flat space-time, they can
be fixed by fitting some observable quantities in the vacuum, for
instance, the pion mass and the pion decay constant. Since we are
not familiar with the particle properties in curved spaces, we are
in principle not able to determine the two parameters in the
Einstein universe. However, what we want to do in this paper is a
general study of the gravitational effect on the chiral phase
structure, without considering a specific system. We focus only on
the qualitative dependence of the phase structure on the gravity,
and do not care very much the precise critical values of the
curvature, temperature, and chemical potential. To this end we scale
the thermodynamic potential (\ref{omega5}) by the soft cutoff
$\Lambda$,
\begin{widetext}
\begin{equation}
\label{omega6}
{\Omega\over
\Lambda^D}=\frac{N(\sigma/\Lambda)^2}{4\Lambda^{D-2}G}-\frac{N}
{\Lambda^{D-1}V}\sum_{l=0}^\infty d_l e^{-{\omega_l\over
\Lambda}}\Bigg[{E_l\over \Lambda}+{T\over \Lambda}
\ln\left(1+e^{-{E_l/\Lambda-\mu/\Lambda\over
{T/\Lambda}}}\right)+{T\over
\Lambda}\ln\left(1+e^{-{E_l/\Lambda+\mu/\Lambda\over
{T/\Lambda}}}\right)\Bigg].
\end{equation}
Without causing confusion, we represent the scaled dimensionless
quantities $\Omega/\Lambda^D,\ \sigma/\Lambda,\ \Lambda^{D-2}G,\
\Lambda^{D-1}V$, $R/\Lambda^2,\ T/\Lambda,\ \mu/\Lambda,\
\omega_l/\Lambda$ still by the corresponding quantities $\Omega,\
\sigma,\ G,\ V,\ R,\ T,\ \mu,\ \omega_l$, and rewrite
(\ref{omega6}) in the dimensionless form,
\begin{equation}
\label{omega7}
\Omega=\frac{N\sigma^2}{4G}-\frac{N}{V}\sum_{l=0}^\infty d_l
e^{-\omega_l}\Big[E_l+T\ln\left(1+e^{-{E_l-\mu\over T}}\right)
+T\ln\left(1+e^{-{E_l+\mu\over T}}\right)\Big].
\end{equation}
\end{widetext}
Now only one dimensionless parameter, the coupling constant $G$
appears in the model.

Calculating the first order derivative of the thermodynamic
potential (\ref{omega7}) with respect to the condensate $\sigma$,
we obtain the gap equation which determines the $T-$, $\mu-$ and
$R-$dependence of the condensate,
\begin{equation}
\label{gap6}
\sigma\left[1-2G I(T,\mu,R,\sigma)\right]=0,
\end{equation}
with the function $I$ defined as
\begin{equation}
\label{i}
I = {1\over V}\sum_{l=0}^\infty {d_l\over
E_l}e^{-\omega_l}\left(1-f(E_l+\mu)-f(E_l-\mu)\right),
\end{equation}
where $f(x)=1/(e^{x/T}+1)$ is the Fermi-Dirac distribution
function. The trivial solution $\sigma=0$ of the gap equation
describes the symmetry restoration phase, and the other solution
$\sigma\ne 0$ corresponds to the symmetry breaking phase. It is
necessary to note that the solution of the gap equation is not
guaranteed to be the physical condensate, we should check for each
solution if it is the position of the minimum thermodynamic
potential.

What happens when the curvature tends to zero? While the line
element (\ref{einstein}) of the Einstein space-time can not return
to the line element of Minkowskian space-time when $R\rightarrow
0$ or $a\rightarrow\infty$, because the two space-times have
different topologies, the thermodynamic potential in the limit
$R\rightarrow 0$ becomes the one in the flat space-time. In the
limit of $a\rightarrow\infty$, we have
\begin{eqnarray}
\label{limit}
&& \omega_l\rightarrow a^{-1} l,\ \ \ \ \ d\omega_l=\omega_{l+1}-\omega_l\rightarrow a^{-1},\nonumber\\
&& d_l\rightarrow
\frac{2^{[(D+1)/2]}a^{D-2}\omega_l^{D-2}}{(D-2)!},
\end{eqnarray}
and the thermodynamic potential is reduced to
\begin{widetext}
\begin{equation}
\label{flat}
\Omega\rightarrow\frac{N\sigma^2}{4G}-\frac{N2^{[(D-1)/2]}\Gamma(D/2)}{\pi^{D/2}\Gamma(D-1)}\int_0^\infty
d p p^{D-2}e^{-p} \left[E_p+T\ln\left(1+e^{-{E_p-\mu\over
T}}\right)+T\ln\left(1+e^{-{E_p+\mu\over T}}\right)\right],
\end{equation}
\end{widetext}
where we have replaced the integrated variable $\omega_l$ by $p$
which could be considered as the fermion momentum. This is exactly
the familiar formula for the thermodynamic potential of the NJL
model in $D$-dimensional Minkowskian space-time.

\begin{figure}[!htb]
\begin{center}
\includegraphics[width=6cm]{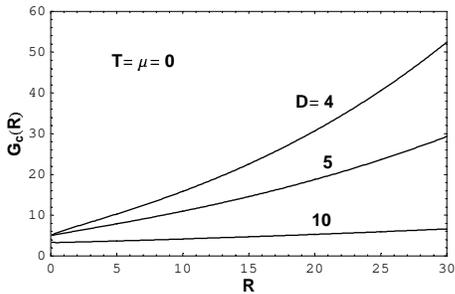}
\caption{The critical coupling constant $G_c$ as a function of $R$
for three values of $D$. At each value of $D$ the chiral breaking
phase corresponds to the region above the line. } \label{fig1}
\end{center}
\end{figure}

Before discussing the chiral properties at finite temperature and
density, we first consider the pure gravitational effect. In the
curved vacuum with $T=\mu=0$, the thermodynamic potential is reduced
to
\begin{equation}
\label{potential}
\Omega=\frac{N\sigma^2}{4G}-\frac{N}{V}\sum_{l=0}^\infty
d_lE_le^{-\omega_l},
\end{equation}
and correspondingly, the gap equation becomes
\begin{equation}
\label{gap7}
 0=\sigma\left[1-\frac{2G}{V}\sum_{l=0}^\infty
\frac{d_l}{E_l}e^{-\omega_l}\right].
\end{equation}
To have chiral symmetry breaking $\sigma\ne 0$ in the vacuum, the
coupling constant $G$ must exceed the critical value $G_c$ which
is determined by
\begin{equation}
\label{gc}
1=\frac{2G_c}{V}\sum_{l=0}^\infty
\frac{d_l}{\omega_l}e^{-\omega_l}.
\end{equation}
The critical coupling constant $G_c$ as a function of the curvature
$R$ is shown in Fig.\ref{fig1} for the dimension $D=4, 5$ and $10$.
As $R$ increases, $G_c$ growths monotonously, reflecting the fact
that the broken chiral symmetry can be restored by finite curvature
effect. To keep chiral symmetry breaking in curved vacuum with
finite $R$, it needs stronger interaction than that in flat vacuum.
However, this curvature effect is gradually washed out when the
dimension of the space-time increases. From Fig.\ref{fig1}, the
effect is almost fully cancelled in the case of $D=10$. To have a
remarkable curvature effect, we consider in the following only the
cases with $D=4$ and $5$. The initial value of $G_c$ at $R=0$ can be
analytically derived by taking the limit $a\rightarrow \infty$ in
Eq.(\ref{gc}),
\begin{equation}
\label{g0}
G_c(R=0)={(D-2)\pi^{D/2}2^{-[(D+1)/2]}\over
\Gamma(D/2)}.
\end{equation}
\begin{figure}[!htb]
\begin{center}
\includegraphics[width=6cm]{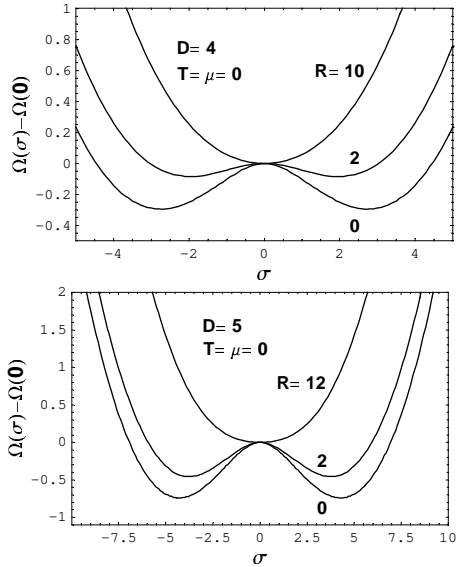}
\caption{The shifted thermodynamic potential as a function of
chiral condensate at $T=\mu=0$ for $R=0, 2, 10$ at $D=4$ and $R=0,
2, 12$ at $D=5$ } \label{fig2}
\end{center}
\end{figure}

\section {Phase Diagrams}
\label{s4}
To guarantee chiral symmetry breaking in the vacuum, we take the
coupling constant $G=10>G_c(0)$ for $D=4$ and $5$ in the following
numerical calculations.

The curvature effect on the chiral symmetry restoration can be
clearly seen in the $R-$dependence of the thermodynamic potential.
In Fig.\ref{fig2} we plot the shifted thermodynamic potential in
the curved vacuum, $\Omega(\sigma)-\Omega(0)$, as a function of
$\sigma$ for different values of $R$. At small values of $R$, the
physical condensate corresponding to the minimum thermodynamic
potential is finite but not zero, which means chiral symmetry
breaking, but at large values of $R$, the minimum of the
thermodynamic potential is located at $\sigma=0$, which stands for
chiral symmetry restoration. It is easy to see from
Figs.\ref{fig1} and \ref{fig2} that the critical curvature $R_c$
for chiral restoration increases with increasing dimension.

We now calculate the chiral condensate at finite temperature,
chemical potential and curvature. At $T=0$, the gap equation
(\ref{gap6}) is reduced to
\begin{equation}
\label{gap7}
\sigma\left[1-2\frac{G}{V}\sum_{l=0}^\infty\frac{d_l}{E_l}e^{-\omega_l}\theta(E_l-\mu)
\right]=0,
\end{equation}
where $\theta(x)$ is the step function. Since $E_l$ grows
monotonously with increasing $l$, the nonzero chiral condensate is
just a constant $\sigma_0$ in the region
\begin{equation}
\label{uc}
\mu\le\mu_0=E_0=\sqrt{\omega_0^2+\sigma_0^2}
\end{equation}
with $\sigma_0$ determined by
\begin{equation}
\label{s0}
1-2\frac{G}{V}\sum_{l=0}^\infty\frac{d_l}{E_l}e^{-\omega_l}=0.
\end{equation}
However, as we mentioned above, the solution $\mu_0$ of the gap
equation is probably not the critical chemical potential $\mu_c$
for chiral phase transition. We should check the minimum of the
thermodynamic potential
\begin{equation}
\label{omega0}
\Omega(\mu,R,\sigma)={\sigma^2\over 4G}-{1\over
V}\sum_{l=0}^\infty d_l
e^{-\omega_l}\left[\mu+\left(E_l-\mu\right)\theta(E_l-\mu)\right],
\end{equation}
and determine $\mu_c$ by the condition
\begin{equation}
\label{muc}
\Omega(\mu_c,R,\sigma_0)=\Omega(\mu_c,R,0).
\end{equation}
At $R=0$, we have $\mu_c=2.2<\mu_0=2.9$ for $D=4$ and
$\mu_c=3.5<\mu_0=4.7$ for $D=5$. At fixed $R$, the physical
condensate $\sigma$ is simply a step function of $\mu$,
\begin{equation}
\label{step}
\sigma(\mu)=\sigma_0\theta(\mu_c-\mu).
\end{equation}

\begin{figure}[!htb]
\begin{center}
\includegraphics[width=6cm]{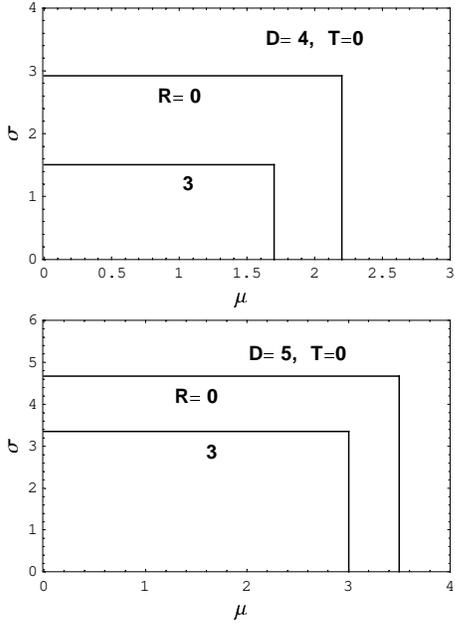}
\caption{The chiral condensate $\sigma$ as a function of chemical
potential $\mu$ at $T=0$ and for $R=0, 3$ and $D=4, 5$.}
\label{fig3}
\end{center}
\end{figure}

We show the chemical potential dependence of $\sigma$ in
Fig.\ref{fig3} at $T=0$ and for $R=0$ and $3$. The critical
chemical potential drops down with increasing curvature and
finally reaches $\sqrt{\omega_0^2(R_c)+\sigma_0^2(R_c)}$ at the
critical curvature $R_c$. It is easy to see that $R_c$ satisfies
the condition
\begin{equation}
\label{rc}
1-2\frac{G}{V}\sum_{l=0}^\infty\frac{d_l}{\omega_l}e^{-\omega_l}=0.
\end{equation}
At $D=4$ and $5$, the critical curvature $R_c$ are, respectively,
4.7 and 8.5. Above $R_c$ there exists no more chiral breaking
phase and the condensate keeps zero at any chemical potential.

\begin{figure}[!htb]
\begin{center}
\includegraphics[width=7cm]{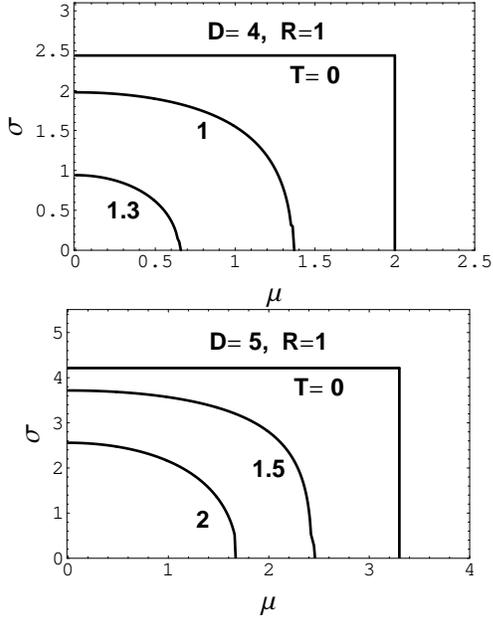}
\caption{The chiral condensate $\sigma$ as a function of chemical
potential $\mu$ at fixed curvature $R=1$ and for $T=0, 1, 1.3$ at
$D=4$ and $T=0, 1.5, 2$ at $D=5$.} \label{fig4}
\end{center}
\end{figure}

The chiral condensate at finite temperature is shown in
Fig.\ref{fig4} at fixed $R$. Similar to the case in flat
space-time, the temperature effect leads to chiral symmetry
restoration. When $\mu=0$, the chiral condensate decreases with
increasing temperature and finally reaches zero at the critical
value $T_c$ determined by
\begin{equation}
\label{tc}
1-2\frac{G}{V}\sum_{l=0}^\infty\frac{d_l}{\omega_l}e^{-\omega_l}\left(1-2f(\omega_l)\right)=0.
\end{equation}
Above $T_c$ the chiral breaking phase disappears at any chemical
potential.
\begin{figure}[!htb]
\begin{center}
\includegraphics[width=7.5cm]{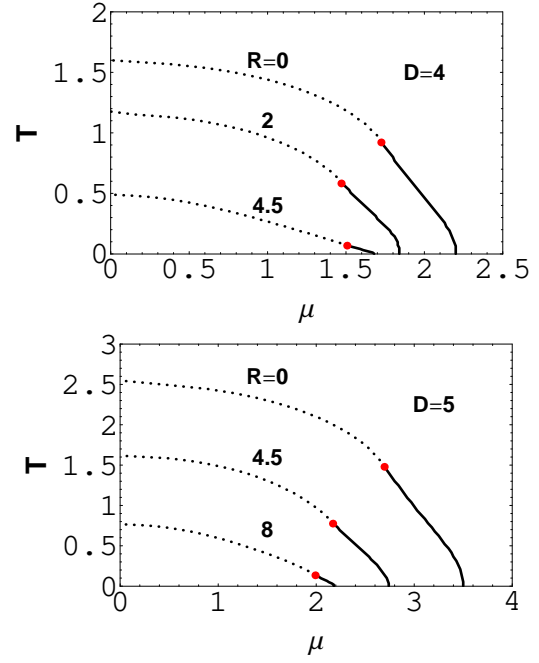}
\caption{The phase diagram in $T-\mu$ plane for $R=0, 2, 4.5$ at
$D=4$ and $R=0, 4.5, 8$ at $D=5$. The dashed and solid lines
represent, respectively, the second and first order phase
transitions, and the dots indicate the tricritical points which
connect the corresponding second and first order transitions. }
\label{fig5}
\end{center}
\end{figure}
\begin{figure}[!htb]
\begin{center}
\includegraphics[width=7.5cm]{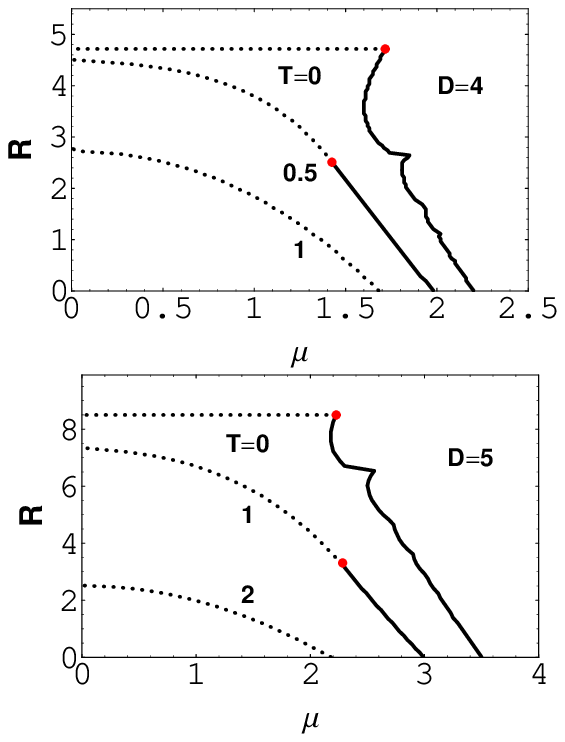}
\caption{The phase diagram in $R-\mu$ plane for $T=0, 0.5, 1$ at
$D=4$ and $T=0, 1, 2$ at $D=5$. The dashed and solid lines
represent, respectively, the second and first order phase
transitions, and the dots indicate the tricritical points.}
\label{fig6}
\end{center}
\end{figure}
\begin{figure}[!htb]
\begin{center}
\includegraphics[width=7.5cm]{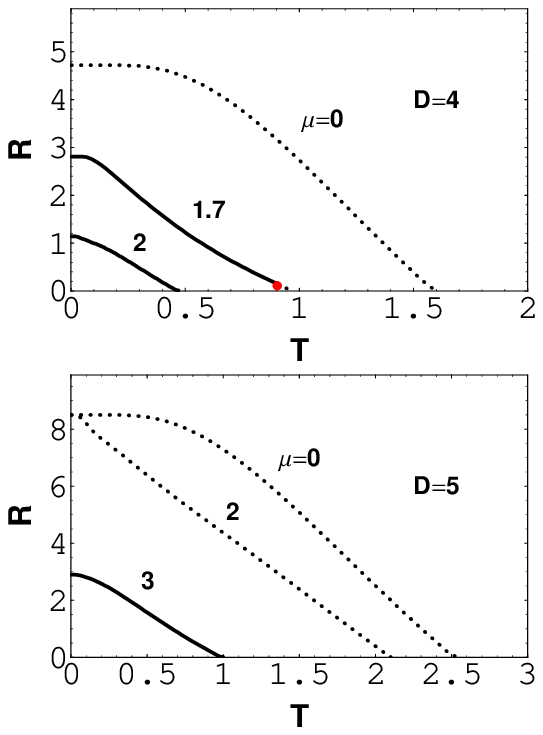}
\caption{The phase diagram in $R-T$ plane for $\mu=0, 1.7, 2$ at
$D=4$ and $\mu=0, 2, 3$ at $D=5$. The dashed and solid lines
represent, respectively, the second and first order phase
transitions, and the dots indicate the tricritical points.}
\label{fig7}
\end{center}
\end{figure}

The phase diagram in the $T-\mu$ plane at fixed $R$ is shown in
Fig.\ref{fig5}. It is very similar to the familiar phase structure
in flat space-time. Each phase transition line separates the
chiral breaking phase below the line and the chiral restoration
phase above the line. The chiral breaking region is gradually
suppressed with increasing curvature and finally disappears at the
critical value $R_c$. The dashed and solid lines represent,
respectively, the second and first order chiral phase transitions,
and the dots indicate the tricritical points which link the
corresponding second and first order transitions.

The phase diagrams in $R-\mu$ and $R-T$ planes are shown,
respectively, in Figs.\ref{fig6} and \ref{fig7}. Again the dashed
and solid lines mean the second and first order phase transitions,
and the dots indicate the tricritical points. Just as what we
expected, at fixed curvature, the phase transition becomes more and
more easy when the temperature or chemical potential increases. The
straight dashed lines in the $R-\mu$ plane at $T=0$ reflect the fact
that the Einstein universe is compact and in turn the lowest
eigenvalue $\omega_0$ of equation (\ref{eigen2}) is not zero. The
toothed lines in Fig.\ref{fig6} at $T=0$ come from the step function
in equation (\ref{gap7}) which makes the low limit of the summation
jump up when $\mu$ increases. From the phase diagrams in
Figs.\ref{fig5}, \ref{fig6} and \ref{fig7}, the second order phase
transition happens at high $T$ or high $R$ or low $\mu$, while the
first order phase transition occurs at low $T$ or low $R$ or high
$\mu$. While finite values of $R, T$ and $\mu$ can all lead to a
chiral phase transition, the order behavior of the phase transition
induced by curvature effect is similar to that by temperature
effect, but different from that by chemical potential effect.
Therefore, in the sense of chiral restoration, the gravitational
effect is more like the temperature effect.

We now discuss the critical properties for the second order chiral
phase transition. Expanding the gap equation for $\sigma$ in the
chiral breaking phase
\begin{equation}
\label{c1}
1-2GI(T,\mu,R,\sigma)=0
\end{equation}
around the critical value $R_c$ at fixed $T$ and $\mu$,
\begin{equation}
\label{c2}
a{R-R_c\over R_c}+b\sigma^2+c\sigma^4=0
\end{equation}
with the coefficients $a, b$ and $c$ defined as
\begin{eqnarray}
\label{c3} a(T,\mu,R_c)&=&R_c{\partial I\over \partial
R}\Big|_{\sigma=0},\nonumber\\
b(T,\mu,R_c)&=&{\partial I\over \partial
\sigma^2}\Big|_{\sigma=0},\nonumber\\
c(T,\mu,R_c)&=&{1\over 2}{\partial^2 I\over \partial
(\sigma^2)^2}\Big|_{\sigma=0}
\end{eqnarray}
at the critical point, where we have considered the fact that $I$
is a function of $\sigma^2$ and neglected the higher orders of
$\sigma^2$.

The critical condensate $\sigma_c$ is then determined in the limit
$R\rightarrow R_c$,
\begin{equation}
\label{c4}
\sigma_c^2\left(b+c\sigma_c^2\right)=0.
\end{equation}
For a second order phase transition, there should be only one
solution $\sigma_c=0$, the coefficient $c$ must vanish. Therefore,
around the critical point, $\sigma$ behaves as
\begin{equation}
\label{c5}
\sigma(T,\mu,R)=\sqrt{a\over b}\left|{R-R_c\over
R_c}\right|^{1/2},
\end{equation}
which means the critical exponent $\beta_R=1/2$.

For a first order phase transition, however, there should be two
values of $\sigma_c$ at the critical point, one is $\sigma_c=0$,
and the other is $\sigma_c\ne 0$. From (\ref{c4}), the nonzero
condensate is $\sigma_c=\sqrt{-b/c}$. Approaching to the
tricritical point which is the end of the first order phase
transition and the beginning point of the second order phase
transition, the nonzero $\sigma_c\rightarrow 0$, we have therefore
\begin{equation}
\label{c6}
b(T,\mu,R_c)=0
\end{equation}
at the tricritical point. In this case, $\sigma$ around the
tricritical point behaves as
\begin{equation}
\label{c7}
\sigma(T,\mu,R)=\sqrt[4]{a\over c}\left|{R-R_c\over
R_c}\right|^{1/4},
\end{equation}
which means $\beta_R=1/4$ at the tricritical point. Obviously, the
discussion above for curvature is also valid for temperature, we
have $\beta_T=1/2$ for the second order phase transition and
$\beta_T=1/4$ at the tricritical point.

\begin{figure}[!htb]
\begin{center}
\includegraphics[width=6.5cm]{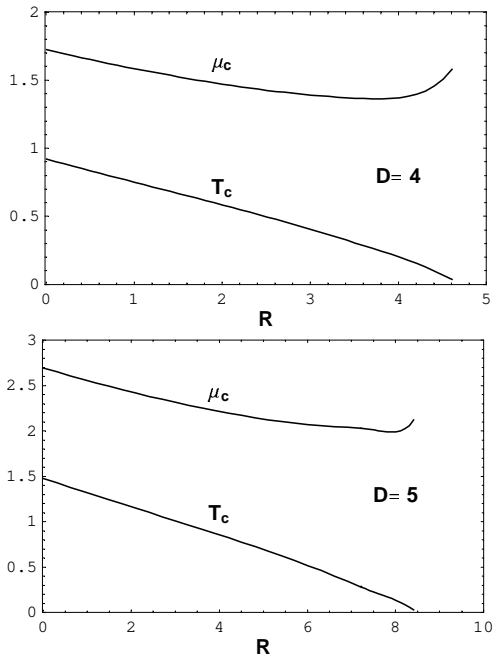}
\caption{The tricritical temperature and chemical potential as
functions of the curvature for $D=4$ and $5$.} \label{fig8}
\end{center}
\end{figure}

The location $(T_c,\mu_c)$ of the tricritical point at fixed $R$
is determined through its definition (\ref{c6}) together with the
gap equation,
\begin{eqnarray}
\label{c8}
&& 1-2GI(T_c,\mu_c,R,0)=0,\nonumber\\
&& b(T_c,\mu_c,R)=0.
\end{eqnarray}
Fig.\ref{fig8} shows the tricritical $T_c$ and $\mu_c$ as
functions of $R$. $T_c$ ends at zero, while $\mu_c$ ends at
$\sqrt{\omega_0^2(R_c)+\sigma_0^2(R_c)}$.

\section {Summary}
\label{s5}
We have investigated the gravitational effects on the chiral
properties at finite temperature and density in the framework of the
NJL model in the Einstein universe. In the mean field approximation,
the thermodynamic potential and the gap equation to determine the
chiral condensate are analytically derived without any further
approximation, which allows us to reasonably study the chiral
symmetry around the phase transition where any perturbative
expansion is in principle not valid.

If we take the curvature $R$ of the space-time which describes the
gravitational effect in the Einstein universe as an external
parameter, the role it plays in the chiral phase transition is very
much like the effect of temperature. Both finite $R$ and $T$ result
in chiral phase transitions and the behavior of the order of the
transition is similar. However, this gravitational effect on the
phase transition will be washed out when the dimension of the
universe is high enough.

The chiral phase diagram in the $T-\mu$ plane is similar to the one
in flat space-time where the thermodynamics can be obtained from
that in the Einstein universe by taking the limit $R\rightarrow 0$.
However, the compact property of the Einstein universe makes the
minimum critical chemical potential $\mu_c$ at $T=0$ be finite but
not zero, which leads to a flat roof structure in the phase diagram
in the $R-\mu$ plane at $T=0$. From the definition of the
tricritical point, we determined its location in the $T-\mu$,
$R-\mu$ and $R-T$ planes.

It should be noted here that in the problem of the early universe
the effects of external gauge field on the chiral phase transition
are important. These effects have been studied in \cite{geye, geye1,
geye2}, and our formulation permits the inclusion of these effects.

The change in the chiral properties induced by finite curvature
effects may be useful for the investigation of compact stars where
the gravitational effects can not be neglected. A natural extension
of the study is to discuss the di-fermion condensate in curved
space-time, which is significant for the study of color
superconductivity, since it may exist in the core of compact stars.

{\bf Acknowledgments:} X.G.H thanks professor T.Inagaki for
reminding us the references \cite{inag4,ishi}. The work was
supported by the grants NSFC10575058, 10435080, 10428510, and
SRFDP20040003103.

\end{document}